\documentstyle{article}

\newcommand{\be}{\begin{equation}}
\newcommand{\ee}{\end{equation}}
\newcommand{\bea}{\begin{eqnarray}}
\newcommand{\eea}{\end{eqnarray}}

\newcommand{\e}{{\rm e}}

\newcommand{\ie}{{\it i.e.,\ }}

\begin{document}
\begin{titlepage}
\addtolength{\baselineskip}{0.20\baselineskip}
\begin{tabbing}
\ \hskip 12cm \= NDA-FP-12/93 \\
\> OCHA-PP-33 \\
\> INS-Rep.987 \\
\> \ \\
\> June 1993
\end{tabbing}

%\vfill

\begin{center}
\Large
{\sc Black Hole Physics from Two Dimensional Dilaton Gravity based on
$SL(2,R)/U(1)$ Coset Model}

\vfill

\normalsize
%\vspace{.5cm}
\large
{\sc Shin'ichi Nojiri}

{\it Department of Mathematics and Physics
\\ National Defense Academy
\\ Hashirimizu, Yokosuka 239, JAPAN}

\vfill

%\vspace{.5cm}
\large
{\sc Ichiro Oda}

{\it Institute for Nuclear Study
\\ University of Tokyo
\\ 3-2-1 Midori-cho, Tanashi-shi, Tokyo 188, JAPAN}

{\rm and}

{\it Faculty of Science, Department of Physics
\\ Ochanomizu University
\\ 1-1, Otsuka 2, Bunkyo-ku, Tokyo 112, JAPAN}

\vfill

PACS number(s) : 04.60.+n, 11.17.+y, 97.60.Lf

\end{center}

%\end{document}

%\vfill

\newpage

\begin{abstract}
We  analyze  quantum  two  dimensional dilaton gravity model, which is
described by $SL(2,R)/U(1)$ gauged Wess-Zumino-Witten model deformed by
$(1,1)$  operator.  We  show  that  the  curvature singularity does not
appear when the central charge $c_{\rm matter}$ of the matter fields is
given  by  $22<c_{\rm matter}<24$.   When  $22<c_{\rm matter}<24$,  the
matter  shock  waves,  whose  energy  momentum  tensors  are  given  by
$T_{\rm matter} \propto \delta(x^+ - x^+_0)$, create a kind of wormholes,
\ie causally disconnected regions. Most of the quantum informations in
past  null  infinity  are  lost  in  future  null infinity but the lost
informations would be carried by the wormholes.

We  also  discuss  about  the problem of defining the mass of quantum
black holes.  On the basis of the argument by Regge and Teitelboim,
we show that the ADM mass  measured  by  the  observer  who  lives in one
of asymptotically flat regions is finite and does not vanish in general.
On the other hand, the Bondi mass is ill-defined in this model.
Instead of the Bondi mass, we consider the mass measured by observers who
live in an asymptotically flat region at first. A class of the observers
finds the mass of the black hole created by a shock wave changes as the
observers' proper time goes by, \ie they observe the Hawking radiation.
The measured mass vanishes after the infinite proper time and
the black hole evaporates completely.
Therefore the total Hawking radiation is positive even when $N<24$.

\end{abstract}
\end{titlepage}

%\end{document}

\newpage

\section{Introduction}

When we try to construct  the quantum theory of the gravity,  black hole
evaporation provides a serious problem \cite{O}.  A toy model proposed
by Callan, Giddings, Harvey and Strominger (CGHS) \cite{I} was expected
to  give  a  clue  to  solve  this  problem.  The  model  describes  the
two-dimensional  gravity  coupled  with a dilaton and conformal matters.
Many authors have investigated this model \cite{IA}--\cite{CIII} and
it has been pointed out in several papers \cite{II,III,AI} that
the  consistent  quantization  of  this theory  in  the  conformal gauge
\be
\label{ei}
g_{\mp\pm}=-{1 \over 2}\e^{2\rho}\ ,\ \ g_{\pm\pm}=0 \ ,
\ee
requires  that  the theory  should  be  a conformal field theory with the
vanishing central charge. The quantum action of the theory is given by a
sum of the actions of two free fields and the dilatonic cosmological term,
which is a $(1,1)$--operator \cite{II,III}. Recently the authors have
proposed  a  new  class of quantum dilaton gravity which is described by
$SL(2,R)/U(1)$  gauged  Wess-Zumino-Witten  model \cite{VI}  deformed
by  $(1,1)$  operator  \cite{CI}.  In  the  weak  coupling   limit
: $\e^{2\phi}\rightarrow 0$  ($\phi$  is  a  dilaton field) the action
reduces to CGHS' classical action.
The model  was analyzed by $1/k$ expansion ($k$ is the level of
$SL(2,R)$  Wess-Zumino-Witten  model  and   we   found   the   curvature
singularity does not appear when $k$ is negative and $|k|$ is large.

In this paper,  we analyze the previously proposed model \cite{CI} in
detail.\footnote{
This paper is the complete and revised version of \cite{CI}.}
We  show  that the curvature singularity does not appear when the
central  charge  $c_{\rm matter}$  of  the  matter  fields  is  given by
$22<c_{\rm matter}<24$. When $22<c_{\rm matter}<24$, the matter shock waves,
whose energy momentum tensor is given by
$T_{\rm matter}\propto \delta(x^+ - x^+_0)$,
create a kind of wormholes, \ie causally disconnected regions.
Most of the quantum informations in past null infinity are lost in future
null infinity but the lost informations are carried by the wormholes.

We also discuss about the problem of defining the mass of quantum black
holes. Recently Bilal, Kogan \cite{CII} and de Alwis \cite{CIII} have
proposed a definition of the mass based on the argument by Regge and
Teitelboim \cite{CIV} and they have found the ADM mass of the black
holes vanishes.
In spite of their interesting observations, we claim that problems are
remained in order to obtain a well-defined mass and the ADM mass measured by
the observer who lives in one of asymptotically flat regions does not vanish
in general.

We also show that the Bondi mass is ill-defined in this model since
the scalar curvature diverges at a point in the infinte future.
This curvature singularity, however, would not give any other serious
problem.
Instead of the Bondi mass, we consider the mass measured by observers who
live in an asymptotically flat region at first. A class of the observers
finds the mass of the black hole created by a shock wave changes as the
observers' proper time goes by, \ie they observe the Hawking radiation.
The measured mass vanishes after the infinite proper time and
the black hole evaporates completely.
Therefore the total Hawking radiation is positive even when $N<24$.

In the next section, we review the model proposed in the previous
paper \cite{CI}. We give the action of the model and solve the equations
of motion explicitly. By using the solutions, we discuss the structure of
the space-time in section 3. Especially, we consider the problem of the
loss of the quantum informations. In section 4, we define the mass of the
quantum black holes. Last section is devoted to summary and discussion.

\section{A Model of Quantum Dilaton Gravity based
on $SL(2,R)/U(1)$ Coset Model}

In this section, we review the model proposed previously \cite{CI}.

In  the  original  paper  by  CGHS \cite{I}, the quantum effects were
expected  to  be  described by  adding correction term which only comes
from  the conformal anomaly to the classical action.  It was,  however,
clarified  \cite{II,III,AI}  that we need more counterterms since the
quantum  action  should  have  conformal  symmetry  when  we choose the
conformal gauge  (\ref{ei})  \cite{VA}.  By following de Alwis' paper
\cite{V},  we  assume  the  kinetic  term  of the quantum action in the
dilaton gravity coupled to $N$ free bosons is given by
\begin{eqnarray}
\label{eii}
S_{\rm kin}&=&{1 \over 2\pi}\int d^2x\, \Bigl\lbrack
-8{\rm {\rm e}}^{-2\phi}(1+h(\phi))\partial_+\phi \partial_-\phi \\
&\ &+4{\rm {\rm e}}^{-2\phi}(1+\bar h(\phi))(\partial_+\phi \partial_-\rho
+\partial_+\rho \partial_-\phi)
+2\kappa(1+\bar{\bar h}(\phi))\partial_+\rho \partial_-\rho\Bigr\rbrack\ .
\nonumber
\end{eqnarray}
Here $h(\phi)$,  $\bar h(\phi)$  and $\bar{\bar h}(\phi)$ are ${\cal O}
({\rm e}^{2\phi})$    and    $\kappa={24-N \over 6}$.    Note      that
${\rm e}^{2\phi}$  plays  the role of space-time dependent gravitational
coupling \cite{I}. In Ref.\cite{V}, it was only considered the case
$\bar{\bar h}(\phi)=0$,  where  $S_{\rm kin}$  can  be rewritten by two
free fields action.  In Ref.\cite{CI}, the authors considered
$\bar{\bar h}(\phi)\neq 0$ case.

If we define new fields $X$ and $Y$ by,
\begin{eqnarray}
\label{eiii}
X&\equiv&\pm 2b \int^\phi ds {\rm {\rm e}}^{-2s}
\sqrt{(1+\bar h(s))^2(1+\bar{\bar h}(s))^{-1}+\kappa
{\rm {\rm e}}^{2s}(1+h(s))} \ , \nonumber \\
Y&\equiv&\pm a \Bigl\lbrack\rho
+{2 \over \kappa}\int^\phi ds {\rm {\rm e}}^{-2s}
(1+\bar h(s))(1+\bar{\bar h}(s))^{-1} \Bigr\rbrack\ ,
\end{eqnarray}
the kinetic term of the quantum action  $S_{\rm kin}$ in Eq.(\ref{eii})
can be rewritten by,
\begin{equation}
\label{eiv}
S_{\rm kin}={k \over 4\pi}\int d^2x\, \Bigl\lbrack \partial_+ X\partial_- X
-\Bigl(1+\bar{\bar h}(\phi(X))\Bigr)
\partial_+ Y \partial_- Y \Bigr\rbrack\ .
\end{equation}
Here $a$ and $b$ are defined by
\begin{equation}
\label{eiva}
a\equiv\sqrt{-{4\kappa \over k}}\ ,\ \ \ b\equiv\sqrt{-{4 \over k\kappa}}\ ,
\end{equation}
and $k$ is a constant satisfying  $k\kappa <0$ and we will identify $k$ with
the level of $SL(2,R)$ Wess-Zumino-Witten model later. If $\bar{\bar h}
(\phi)$ in the action (\ref{eiv}) is a constant,  this action describes
a free field theory,  which is the simplest conformal field theory.  On
the  other hand,  if we can choose $1+\bar{\bar h}(\phi(X))=\tanh^2 X$,
the  action  (\ref{eiv}) is nothing but the action of another conformal
field theory,  {\it i.e.,\ }  $SL(2,R)/U(1)$  gauged Wess-Zumino-Witten
model in unitary gauge \cite{VII}.
\begin{equation}
\label{eix}
S_{\rm kin}={k \over 4\pi}\int d^2x\, \Bigl\lbrack \partial_+ X\partial_- X
-\tanh^2 X \partial_+ Y \partial_- Y \Bigr\rbrack\ .
\end{equation}
In fact, if we choose, for example,
\begin{eqnarray}
\label{eaii}
1+h(\phi)&=&-{{\rm e}^{-2\phi} \over \kappa}
\Bigl\{\Bigl(1-{\kappa{\rm e}^{2\phi} \over 2}\Bigr)^2
\tanh^2(b{\rm e}^{-2\phi})-1\Bigr\} \ , \nonumber \\
1+\bar h(\phi)&=&\Bigl(1-{\kappa{\rm e}^{2\phi} \over 2}\Bigr)
\tanh^2(b{\rm e}^{-2\phi}) \ , \nonumber \\
1+\bar{\bar h}(\phi)&=&\tanh^2(b{\rm e}^{-2\phi}) \ ,
\end{eqnarray}
{\it i.e.,\ } when $X$ and $Y$ are given by,
\begin{equation}
\label{eaiii}
X=b{\rm e}^{-2\phi}\ , \ \ \
Y=a\Bigl(\rho-\phi-{1 \over \kappa}{\rm e}^{-2\phi}\Bigr)\ ,
\end{equation}
we find the action (\ref{eiv}) is rewritten in the form of Eq.(\ref{eix}).
Note that there appears non-perturbative contribution of
${\cal O}({\rm {\rm e}}^{-{2b \over {\rm {\rm e}}^{2\phi}}})$.

The cosmological term which is $(1,1)$-operator can be added to the action
(\ref{eix}).
The vertex operator $V_{lm}$, which is an $(l,m)$ representation of
$SL(2,R)$, has the following form in the unitary gauge \cite{VIII},
\begin{equation}
\label{exi}
V_{lm}=(\sinh^2 X)^l{\rm {\rm e}}^{-2mY}F(m-l, -m-l, 1 ; \coth^2 X)\ .
\end{equation}
Here $F(\alpha , \beta , \gamma ; x)$ is Gauss' hypergeometric function.
If we require
$V_{lm}\sim {\rm {\rm e}}^{c(X+Y)} \sim {\rm {\rm e}}^{c'(\rho-\phi)}$
($c$ and $c'$ are positive constants.) when $X\longrightarrow \infty$
{\it i.e.,\ } ${\rm e}^{2\phi} \longrightarrow 0$ (weak coupling limit),
which is expected from the $\beta$-function analysis \cite{V},
we find $l=- m$ and
\begin{equation}
\label{exiii}
V_l\equiv V_{ll}=(\sinh^2 X)^l{\rm {\rm e}}^{2lY}\ .
\end{equation}
Then the total action including cosmological term should be given by,
\begin{eqnarray}
\label{eixx}
S&=&{k \over 4\pi}\int d^2x\, \Bigl\lbrack \partial_+ X\partial_- X
-\tanh^2 X \partial_+ Y \partial_- Y
+ {\alpha \over 4k}(\sinh X{\rm {\rm e}}^Y)^{2l}\Bigr\rbrack \nonumber \\
&\ &+(N\ {\rm free\ boson\ terms})\ .
\end{eqnarray}
We choose $l$ so that the conformal dimension $\Delta_l$ of the
operator $V_l$ should be a unity : $\Delta_l=1$

In order to consider the stress tensors, we rewrite the action
(\ref{eixx}) in a reparametrization invariant form,
\begin{eqnarray}
\label{exx}
S&=&{k \over 8\pi}\int \sqrt{-g} d^2x\, \Bigl\lbrack -g^{\mu\nu}
\partial_\mu X(\phi) \partial_\nu X(\phi)\nonumber \\
&\ &+\tanh^2 X g^{\mu\nu}
\partial_\mu \Bigl(a\hat\rho+\hat Y(\phi)\Bigr)
\partial_\nu \Bigl(a\hat\rho+\hat Y(\phi)\Bigr) \nonumber \\
&\ &+ {\alpha \over k}\Bigl(\sinh X
{\rm e}^{a\hat\rho+\hat Y(\phi)}\Bigr)^{2l}{\rm e}^{-2\hat\rho}\Bigr\rbrack
+(N\ {\rm free\ boson\ terms})\ .
\end{eqnarray}
Here we have used the parametrization in Eq.(\ref{eaiii}) and we define
$\hat Y(\phi)$ and $\hat\rho$ by ($R$ is a scalar curvature)
\begin{equation}
\label{exxi}
\hat Y\equiv a\Bigl(-\phi-{1 \over \kappa}{\rm e}^{-2\phi}\Bigr)\ , \ \ \
\hat\rho\equiv -{1 \over 2} (\partial_\mu \sqrt{-g} g^{\mu\nu}
\partial_\nu )^{-1}\sqrt{-g} R\ .
\end{equation}
Then we find the stress tensors $T_{\pm\pm}$ have the following forms,
\begin{eqnarray}
\label{exxii}
T_{\pm\pm}&=&k(\partial_\pm X \partial_\pm X
-\tanh^2 X \partial_\pm Y \partial_\pm Y) \nonumber \\
&\ &+{ka \over 2}\partial_\pm\int^{x^\mp}dy^{\mp}
\Bigl\{\partial_- (\tanh^2X\partial_+ Y)
+\partial_+ (\tanh^2X\partial_- Y)\Bigr\} \nonumber \\
&\ &-{\alpha \over 4}(1-al)\partial_\pm
\int^{x^\mp}dy^{\mp}(\sinh X {\rm {\rm e}}^Y)^{2l}
+T_{\pm\pm}^{\rm matter} \ , \nonumber \\
T_{\pm\mp}&=&-{ka \over 2}
\Bigl\{\partial_- (\tanh^2X\partial_+ Y)
+\partial_+ (\tanh^2X\partial_- Y)\Bigr\} \nonumber \\
&\ &-{al\over 4}\alpha(\sinh X {\rm {\rm e}}^Y)^{2l} \ .
\end{eqnarray}
Here $T_{\pm\pm}^{\rm matter}$ is the stress tensor of $N$ scalar fields.
Note that there appear non-local terms in $T_{\pm\pm}$ and it is ambiguous
how to fix the boundary conditions in these terms.
In the following, we consider the solutions corresponding to the following
boundary conditions:
\begin{eqnarray}
\label{exxiii}
T_{\pm\pm}&=&k(\partial_\pm X \partial_\pm X
-\tanh^2 X \partial_\pm Y \partial_\pm Y) \nonumber \\
&\ &+{ka \over 2}\partial_\pm
\Bigl\{\tanh^2X\partial_\pm Y+\int^{x^\mp}_{\alpha^\mp} dy^{\mp}
\partial_\pm(\tanh^2X\partial_\mp Y)\Bigr\} \nonumber \\
&\ &-{\alpha \over 4}(1-al)\partial_\pm
\int^{x^\mp}_{\beta^\mp} dy^{\mp}(\sinh X {\rm {\rm e}}^Y)^{2l}
+T_{\pm\pm}^{\rm matter}
\ .
\end{eqnarray}
The physical quantities like black hole mass {\it etc.} do not depend
on the boundaries $\alpha^\pm$ and $\beta^\pm$ if we use the equations
of motion. As we will see later, the solutions under the above boundary
conditions give the solutions corresponding to CGHS's classical solutions
in the weak coupling limit.

We now solve the equations of motion and constraints of the system.
In order to do this, it is convenient to define new fields $X^\pm$ by,
\begin{equation}
\label{exxvi}
X^\pm=\pm\sinh X{\rm {\rm e}}^{\pm Y} \ .
\end{equation}
Then the action (\ref{eixx}) is rewritten by
\begin{eqnarray}
\label{exxvii}
S&=&{k \over 4\pi}\int d^2x\, \Bigl\lbrack
-{\partial_- X^+ \partial_+ X^- +\partial_- X^- \partial_+ X^+
\over 2(1-X^+ X^-)}+ {\alpha \over 4k}(X^+)^{2l}\Bigr\rbrack \nonumber \\
&\ & +(N\ {\rm free\ boson\ terms})\ .
\end{eqnarray}
and we obtain the following equations of motion
\begin{eqnarray}
\label{exxviia}
0&=&{X^-\partial_- X^+ \partial_+ X^+ \over (1-X^+ X^-)^2}
+ {\partial_- \partial_+ X^+ \over 1-X^+ X^-} \ , \\
\label{exxviib}
0&=&{X^+\partial_- X^- \partial_+ X^- \over (1-X^+ X^-)^2}
+ {\partial_- \partial_+ X^- \over 1-X^+ X^-}
+ {l\alpha \over 2k}(X^+)^{2l-1}
\ .
\end{eqnarray}
Note that $X^+=w$ ($w$ : constant) satisfies the first equation
(\ref{exxviia}). Then we can solve the second equation (\ref{exxviib}),
\begin{equation}
\label{exxix}
X^+=w\ , \ \ \
X^-={1 \over w}\Bigl(1-{\rm {\rm e}}^{-\Lambda x^+x^-+u(x^+,x^-)}\Bigr) \ .
\end{equation}
Here $\Lambda\equiv -{l\alpha \over 2k}(w)^{2l}$ and
$u(x^+,x^-)=u_+(x^+)+u_-(x^-)$.
Although we cannot obtain the general solution to Eqs.(\ref{exxviia}) and
(\ref{exxviib}), it is important that this class of special solutions
(\ref{exxix}) includes the solutions corresponding to the linear dilaton
vacuum, the classical black hole and the shock wave solutions in the weak
coupling region and we can investigate the time-development which
infalling matters lead to.
{}From now on, we only consider the solutions given by Eq.(\ref{exxix}).
By using the solution (\ref{exxix}), we find the stress tensors
in Eqs.(\ref{exxii}), (\ref{exxiii}) have the following forms,
\begin{equation}
\label{exxxi}
T_{\pm\mp}=0 \ , \ \ \ T_{\pm\pm}=
-{ka \over 4}\partial_{\pm}^2u_\pm(x^\pm)
+T_{\pm\pm}^{\rm matter} \ .
\end{equation}
Now let us impose the only a priori restriction \cite{V} in the present
formalism, which is given by,
\begin{equation}
\label{exxxii}
0=T_{\pm\pm}+t_{\pm\pm} \ , \ \
t_{\pm\pm}=-{N \over 24}{1 \over (x^\pm)^2} \ .
\end{equation}
When $T_{\pm\pm}^{\rm matter}=0$, we find the following
solutions\footnote{
There would be an ambiguity in the coefficient of $\ln |x^\pm|$
\cite{V}. In the following, we assume, as a natural choice, the coefficient
should be $-{N \over 24}$. We can straightforwardly extend the results, which
we will obtain in the following, to the cases of the general coefficient but
most of the results do not be changed qualitatively.}
\begin{equation}
\label{exxxiii}
-{ka \over 4}u_\pm(x^\pm)
=a_\pm+b_\pm x^\pm -{N \over 24}\ln |x^\pm| \ .
\end{equation}
If $b_+=b_-=0$, the solution (\ref{exxxiii}) corresponds to a static object.
On the other hand, the solution corresponding to the shock wave which
describes collapsing matters,
$T_{\pm\pm}^{\rm matter}=m\delta(x^+-x^+_0)$, is given by
\begin{eqnarray}
\label{exxxv}
-{ka \over 4}u_+(x^+)
&=&a_++b_+ x^+ -m(x^+-x^+_0)\theta(x^+-x^+_0)
-{N \over 24}\ln |x^+| \ , \nonumber \\
-{ka \over 4}u_-(x^-)
&=&a_-+b_- x^- -{N \over 24}\ln |x^-| \ .
\end{eqnarray}

Since the action (\ref{eixx}) is correct for large $k$ \cite{VII}, it
is convenient to consider the corresponding action in Landau gauge
\cite{VIII} in order to clarify what kind of conformal theory describes
the system. The Landau gauge action corresponding to (\ref{eixx}) is given by
\begin{eqnarray}
\label{exxxxiii}
S&=&{k \over 4\pi}\int d^2x\, \Bigl\lbrack \partial_+ X\partial_- X
-\cosh^2 X \partial_+ Z \partial_- Z
-\sinh^2 X \partial_+ \tilde Y \partial_- \tilde Y \nonumber \\
&\ &-\Bigl(\cosh^2X-{1 \over 2}\Bigr)
(\partial_+ Z \partial_- \tilde Y - \partial_+ \tilde Y \partial_- Z)
+{1 \over 4}\partial_+ \varphi \partial_- \varphi \nonumber \\
&\ & +{\alpha \over 4k}(\sinh X
{\rm e}^{\tilde Y+{\varphi \over 2}})^{2l}\Bigr\rbrack
+({\rm ghost\ terms}) \nonumber \\
&\ &+(N\ {\rm free\ boson\ terms})\ .
\end{eqnarray}
Here $Z$ is a degree of freedom in $SL(2,R)$ Wess-Zumino-Witten model
which is independent of $X$ and $Y$. A real scalar field $\varphi$ comes
from the $U(1)$ gauge fields and $\tilde Y$ is defined by
$\tilde Y=Y-{\varphi \over 2}$. By using new fields $T$ and $S$, we can
rewrite the action (\ref{exxxxiii}) in a reparametrization invariant and
local form,
\begin{eqnarray}
\label{exxxxiv}
S&=& {k \over 8\pi}\int d^2x\,\Bigl\lbrack
\sqrt{-g} \Bigl\{-g^{\mu\nu} \partial_\mu X(\phi) \partial_\nu X(\phi)
-\cosh^2 X g^{\mu\nu} \partial_\mu Z \partial_\nu Z \nonumber \\
&\ &+\sinh^2 X g^{\mu\nu} \partial_\mu \Bigl(aT+\hat{\hat Y}(\phi)\Bigr)
\partial_\nu \Bigl(aT+\hat{\hat Y}(\phi)\Bigr) \nonumber \\
&\ & -\Bigl(\cosh^2X-{1 \over 2}\Bigr)
\epsilon^{\mu\nu}\partial_\mu Z \partial_\nu \Bigl(aT+\hat{\hat Y}(\phi)\Bigr)
+{1 \over 4}g^{\mu\nu} \partial_\mu \varphi \partial_\nu \varphi
\nonumber \\
&\ &+ {\alpha \over 4k}\Bigl(\sinh X
{\rm {\rm e}}^{aT+\hat{\hat Y}(\phi)+{\varphi \over 2}}\Bigr)^{2l}
{\rm {\rm e}}^{-2T} +{1 \over 2} g^{\mu\nu} \partial_\mu S \partial_\nu T
-{1 \over 4}SR \Bigr\} \Bigr\rbrack\nonumber \\
&\ &+(U(1)\ {\rm ghost\ term})+(N\ {\rm free\ boson\ terms})
\ .
\end{eqnarray}
Here $\hat{\hat Y}(\phi)$ is defined by $\hat{\hat Y}\equiv {\varphi \over 2}
+a\Bigl(-\phi-{1 \over \kappa}{\rm e}^{-2\phi}\Bigr)$.
(See Eq.(\ref{exxi}).)
By using the equation of motion which we obtain by the variation of $S$,
we find
$T=-{1 \over 2} (\partial_\mu \sqrt{-g} g^{\mu\nu} \partial_\nu )^{-1}
\sqrt{-g} R$.
Therefore if we integrate $S$ and $T$ fields we obtain
\begin{eqnarray}
\label{exxxxiva}
S&=& {k \over 8\pi}\int d^2x\,\Bigl\lbrack
\sqrt{-g} \Bigl\{-g^{\mu\nu} \partial_\mu X(\phi) \partial_\nu X(\phi)
-\cosh^2 X g^{\mu\nu} \partial_\mu Z \partial_\nu Z \nonumber \\
&+&\sinh^2 X g^{\mu\nu} \partial_\mu (a\hat\rho +\hat{\hat Y}(\phi))
\partial_\nu (a\hat\rho +\hat{\hat Y}(\phi)) \nonumber \\
&-&\Bigl(\cosh^2X-{1 \over 2}\Bigr)
\epsilon^{\mu\nu}\partial_\mu Z \partial_\nu
(a\hat\rho +\hat{\hat Y}(\phi))
+ {1 \over 4}g^{\mu\nu} \partial_\mu \varphi \partial_\nu \varphi
\nonumber \\
&+& {\alpha \over 4k}\Bigl(\sinh X
{\rm e}^{\hat{\hat Y}(\phi)+{\varphi \over 2}}\Bigr)^{2l}
{\rm e}^{(2la-2)\hat\rho} \Bigr\rbrack\nonumber \\
&+&(U(1)\ {\rm ghost\ term})+(N\ {\rm free\ boson\ terms})
\ .
\end{eqnarray}
Here $\hat\rho $ is defined by Eq.(\ref{exxi}) :
$\hat\rho\equiv -{1 \over 2}
(\partial_\mu \sqrt{-g} g^{\mu\nu} \partial_\nu )^{-1}\sqrt{-g} R$.
Note that the kinetic terms of the resulting action contain the product
of two inverse of d'Alembertian
$\partial_\mu \sqrt{-g} g^{\mu\nu} \partial_\nu$ as in Eq.(\ref{exx}),
which is the reason why we need two fields  $S$ and $T$ in order to localize
the action (\ref{exxxxiva}).

In the conformal gauge, this action (\ref{exxxxiv}) has the following form
\begin{eqnarray}
\label{exxxxivb}
S&=&{k \over 4\pi}\int d^2x\, \Bigl\lbrack \partial_+ X\partial_- X
-\cosh^2 X \partial_+ Z \partial_- Z
+\sinh^2 X \partial_+ \bar Y \partial_- \bar Y \nonumber \\
&\ &+\Bigl(\cosh^2X-{1 \over 2}\Bigr)
(\partial_+ Z \partial_- \bar Y - \partial_+ \bar Y \partial_- Z)
+{1 \over 4}\partial_+ \varphi \partial_- \varphi \nonumber \\
&\ &+ {\alpha \over 4k}(\sinh X{\rm {\rm e}}^{\bar Y+{\varphi \over 2}})^{2l}
{\rm {\rm e}}^{-2\bar T}
-{1 \over 4}(\partial_+ S \partial_- \bar T + \partial_+ S \partial_- \bar T)
\Bigr\rbrack\nonumber \\
&\ & +({\rm ghost\ terms})+(N\ {\rm free\ boson\ terms}) \ .
\end{eqnarray}
Here we have redefined new fields $\bar Y$ and $\bar T$ by
$\bar Y\equiv aT+\hat{\hat Y}(\phi)$ and
$\bar T\equiv T-\rho$.
Then we find that the stress tensors are given by
\begin{eqnarray}
\label{exxxx}
T_{\pm\pm}&=&k\Bigl(\partial_\pm X \partial_\pm X
-\cosh^2 X \partial_\pm Z \partial_\pm Z
+\sinh^2 X \partial_\pm \bar Y \partial_\pm \bar Y \nonumber \\
&\ &+ {1 \over 4}\partial_\pm \varphi \partial_\pm \varphi
+{1 \over 2}\partial_\pm S \partial_\pm \bar T
+{1 \over 4}\partial_\pm^2 S \Bigr)
+T_{\pm\pm}^{\rm matter} \ , \nonumber \\
T_{\pm\mp}&=&{\alpha \over 4}
(\sinh X {\rm {\rm e}}^{\bar Y+{\varphi \over 2}})^{2l}
{\rm {\rm e}}^{-2\bar T}
-{k \over 4}\partial_- \partial_+ S \ .
\end{eqnarray}
When $\alpha=0$, the system is a direct product of those of $SL(2,R)$
Wess-Zumino-Witten model, free boson $\varphi$, ghosts, $N$ free bosons
and $S$-$\bar T$ fields which contribute to the central charge by $2$.
Since the total central charge $c^{\rm total}$ does not depend on $\alpha$,
we find the total central charge $c^{\rm total}$ is given by
\bea
\label{exxiva}
c^{\rm total}&=&c^{SL(2,R)/U(1)}+2-26+N \nonumber \\
&=&{3k \over k-2}-1+2-26+N \nonumber \\
&=&{6 \over k-2}-6\kappa +2 \ .
\eea
Here $c^{SL(2,R)/U(1)}$ is the central charge of $SL(2,R)/U(1)$
gauged Wess-Zumino -Witten model with level $k$
($c^{SL(2,R)/U(1)}={3k \over k-2}-1$).
Since the total central charge  $c^{\rm total}$ should vanish, we obtain
\be
\label{exxivb}
\kappa={1 \over 3}+{1 \over k-2} \ .
\ee
Since $k\kappa<0$, we find $k<-1$ or $0<k<2$ {\it i.e.,\ } we find the
restriction for the number of matter fields $N$ {\it i.e.,\ } $22<N<24$ or
$25<N$.

The operator $\bar V_l$ which corresponds to $V_l$ in Eq.(\ref{exiii})
is given by,
\begin{equation}
\label{exxxxvi}
\bar V_l=(\sinh X{\rm {\rm e}}^{\bar Y+{\varphi \over 2}})^{2l}
{\rm {\rm e}}^{-2\bar T} \ .
\end{equation}
Since ${\rm {\rm e}}^{-2\bar T}$ has the conformal dimension ${1 \over 2}$,
$V_l$ has the dimension $\Delta_l$ as follows,
\be
\label{exxivbb}
\Delta_l=-{l(l+1) \over k-2}-{l^2 \over k}-{1 \over 2} \ .
\ee
We can find $l$ by solving the equation $\Delta_l=1$.

Since the equation of the motion, which is obtained by the variation of $S$,
is given by $\partial_+\partial_-\bar T=\partial_+\partial_- (T-\rho)=0$,
we can fix the residual symmetry of the reparametrization invariance by
choosing the the condition $\bar T=0$, by which we select a special coordinate
system. Under this gauge condition, the solutions of the equations of motion
in the Landau gauge corresponding to the solutions (\ref{exxix}),
(\ref{exxxiii}), (\ref{exxxv}) in the unitary
gauge are given by
\bea
\label{eeeii}
X&=&{1 \over 2}\ln (\e^f+\sqrt{\e^{2f}-1}) \ , \nonumber \\
Y&=&\tilde Y+{\varphi \over 2}={1 \over 2}\ln{w^2 \over \e^f -1} \ ,
\nonumber \\
Z&=&0 \ , \nonumber \\
\varphi&=&-f \ , \nonumber \\
S&=&-2af \ .
\eea
Here $f(x^+,x^-)$ is defined by
\be
\label{eeeiii}
f\equiv -\Lambda x^+x^-+u(x^+,x^-)\ .
\ee

By the action (\ref{eixx}) or (\ref{exxxxivb}), we only consider the
quantum correction which comes from the measure with respect to the
reparametrization invariance and the action does not include any other
counterterms.
In this sense, this action is {\it classical}. By using the knowledge of
current algebra, however, we can calculate amplitudes from the action
(\ref{exxxxivb}). We can also calculate $S$-matrix by solving the equations
of motion which can be obtained from the action (\ref{eixx}) or
(\ref{exxxxivb}). Note that we cannot obtain anything from the equations of
motion which are obtained from quantum effective action, which are given in
terms of a sum of one particle irreducible vertices times semi-classical
fields.

\section{The Structure of Space-Time and the Problem of the Loss of
Quantum Informations}

In this section, we analyze the structure of the space-time which is given
by solutions (\ref{eeeii}) and we show that the curvature singularity
does not appear when the central charge  of the matter fields
$c_{\rm matter}=N$ is given  by  $22<c_{\rm matter}<24$. When
$22<c_{\rm matter}<24$, the matter  shock  waves, whose energy momentum
tensors are given by $T_{\rm matter} \propto \delta(x^+ - x^+_0)$,  create
a kind of wormholes, \ie causally disconnected regions. We also discuss
about the problem of the loss of quantum informations and we claim that
most of the quantum informations in past null infinity are lost in
future null infinity but the lost informations would be carried by the
wormholes.

By using the equations (\ref{eaiii}), we obtain
\be
\label{eeeiv}
\rho={1 \over a}Y-{1 \over2}\ln{X \over b}+{1 \over b\kappa}X \ .
\ee
Substituting the solutions in the equations (\ref{eeeii}), we find $\rho$
is given by
\be
\label{eeev}
\rho={1 \over 2a}\ln{w^2 \over \e^f -1}
-{1 \over2}\ln\Bigr({1 \over 2b}\ln (\e^f+\sqrt{\e^{2f}-1})\Bigl)
+{1 \over 2b\kappa}\ln (\e^f+\sqrt{\e^{2f}-1}) \ .
\ee
Here $f$ is defined in the equation (\ref{eeeiii}). Note that $f$ should
be positive since $\rho$ is real.
Since $\rho$ is given by a monotonously decreasing smooth function with
respect to $f$, the scalar curvature
$R\sim {\rm {\rm e}}^{-2 \rho}\partial_- \partial_+ \rho$
can be singular when $f\rightarrow +0$ or $f\rightarrow +\infty$.
The limit $f\rightarrow +\infty$ corresponds to spacially infinite
asymptotically flat region.
When $f\rightarrow +0$, $\rho$ is given by,
\be
\label{eeevi}
\rho=-{1 \over 2}\Bigr({1 \over a} +{1 \over 2}\Bigl)\ln f + {\cal O}(1)
\ .
\ee
Then the scalar curvature behaves like
\begin{equation}
\label{eaaii}
R\sim f^{{1 \over a}+{1 \over 2}}\Bigl\{{\partial_-\partial_+ f \over f}
-{\partial_- f \partial_+ f \over f^2}\Bigr\} \ ,
\end{equation}
when $f\longrightarrow 0$.
Since $f$ is a smooth function with respect to $x^{\pm}$, except on the
trajectory of matter shock waves, $\partial_\pm f$ or
$\partial_+\partial_- f$ does not diverge on the $f=0$ line.
Therefore if ${1 \over a}-{3 \over 2}\geq 0$, there does not appear
curvature singularity.
Because $a$ is given by $a=\sqrt{-{4\kappa \over k}}$ (\ref{eiva}) and
$\kappa={1 \over 3}+{1 \over k-2}$ (\ref{exxivb}), the curvature singularity
does not appear when $k<-1$ ($k\neq -1$ since $\kappa k\neq 0$), \ie
the number of matter fields $N$ is restricted to $22<N<24$
or more generally, the central charge $c_{\rm matter}$ of the matter fields
is given by $22<c_{\rm matter}<24$.

Since we have obtained a model where the curvature singularity does not
appear, it is important to consider the problem of the loss of quantum
informations {\it etc} by using this model.
For this purpose, we assume $22<c_{\rm matter}<24$ in the following.

Since the metric $g_{\mp\pm}=-{1 \over 2}\e^{2\rho}$ diverges
as $f^{-({1 \over a}+{1 \over 2})}$ when $f\rightarrow +0$, it takes infinte
proper time for anyone to approach to the $f=0$ line.
In Fig.1, the structure of the space-time corresponding to a static solution
(\ref{exxxiii}) is depicted. In Fig.1, $p$ and $q$ are defined by
$x^\pm\equiv\pm \e^{{1 \over 2}(p\pm q)}$. The asymptotically flat region
corresponding to $f\rightarrow +\infty$, which is depicted like a cylinder,
is connected with another asymptotically flat region, which corresponds to
$f\rightarrow +0$ and is depicted like a plane.

In the dynamical solution in the equation (\ref{exxxv}),
we define a function $g(x^+)$ so that $x^-=g(x^+)$ satisfies $f(x^+,x^-)=0$.
We also define $x^-_0$ which satisfies the equation $\partial_+g(x^-_0)=0$.
Then we find that it takes infinite proper time if anyone who lives in the
region I where $x^-<g(x^+)$ and $x^->x^-_0$ (depicted in Fig.2) tries to
get out of this region since his world line always crosses the $f=0$ line.
Therefore this region I is causally disconnected with asymtotically Minkowski
region II. In this sense, this region can be regarded as a kind of wormhole.

We now consider the problem of the loss of the quantum informations.
First we consider a solution corresponding to two matter shock waves
whose stress tensor is given by
\be
\label{eeevii}
T^{++}=\mu_1\delta(x^+-x^+_1)+\mu_2\delta(x^+-x^+_2)\ \ , \ \
(x^+_2>x^+_1)
\ee
and another solution corresponding to one shock wave whose stress tensor is
\be
\label{eeeviii}
T^{++}=(\mu_1+\mu_2)\delta\Bigl(x^+
-{\mu_1 x^+_1+ \mu_2 x^+_2 \over \mu_1 +\mu_2}\Bigr)\ .
\ee
These two solutions have a same form when
$x^+>{\rm max}(x^+_2,{\mu_1 x^+_1+ \mu_2 x^+_2 \over \mu_1 +\mu_2})$:
\bea
\label{eeeix}
-{ka \over 4}u(x^+,x^-)&\equiv& -{ka \over 4}(u_+(x^+)+u_-(x^-))
\nonumber \\
&=& -(\mu_1+\mu_2)x^++\mu_1x^+_1+\mu_2x^+_2-{N \over 24}\ln|x^+x^-|\ .
\eea
This tells that we cannot distinguish these two solutions in future null
infinity ($x^+\rightarrow +\infty$, $x^-$ : fixed). If we define $S$-matrix
between future null infinity and past null infinity ($x^+$ : fixed,
$x^-\rightarrow +\infty$, ) by these solutions, we find the $S$-matrix cannot
be unitary since the $S$-matrix does not have inverse.
Therefore the quantum information in the past null infinity is lost
in the future null infinity.
These two kinds of shock waves, however, create different kinds of wormholes.
The two shock waves solution creates two (Fig.3a) or one (Fig.3b) wormhole.
($n$ shock wave solution often creates $n$ wormholes.)
This suggests that if we consider the Fock
space which includes wormholes, we might possibly construct an $S$-matrix
which is unitary, which might tell that we need to quantize space-time.

\section{The Mass of Black Hole and the Hawking Radiation}

In this section, we discuss about the definition of the mass of two
dimensional black holes in the basis of the argument of Regge and Tetelboim
\cite{CIV}.

In four dimensional gravity theory, if we fix the degrees of
general coordinate transformation completely and choose one of coordinate
systems, we can define a Hamiltonian $H$ with respect to the coordinate
system.
This Hamiltonian $H$ generates the translation of the ``time'' $t$ in
this coordinate system.
The ``time'' has not always any relation with the physical (in some sense)
time and the direction of the ``time'' is not always that of any time-like
Killing vector. (It often happenes that there is not any time-like Killing
vector.)
The operator $\epsilon H$ ($\epsilon$ is an infinitely small constant)
generates a ``time''-translation $t \rightarrow t+\epsilon$ but the physical
distance between $t$ and $t+\epsilon$ is, of course, not $\epsilon$ but
$\sqrt{-g_{00}}\epsilon$.
Therefore what kind of ``time'' translation a Hamiltonian generates depends
on the choice of the coordinate system and the Hamiltonian or its value
\ie mass, has not universal meaning. At least in four-dimension, however,
we can define the mass almost uniquely by according to Regge and Teitelboim.

Let's suppose three dimensional space is a sphere and a black hole lies
on its center.
According to Regge and Teitelboim, the Hamiltonian is given by a sum of the
volume integral of the stress tensor and the surface term.
Since the stress tensor vanishes due to the constraints, only the surface
term can contribute to the value of the Hamiltonian, \ie the mass of the
black hole.
Therefore the mass only depends on the value of the fields on the surface
of the sphere \ie the boundary value of the fields.
If the sphere is large enough, the space-time is asymptotically flat and we
can find an asymptotically time-like Killing vector even if the black hole is
not static.
Since the mass only depends on the boundary value, the mass is independent
of the choice of the coordinate system inside the shell.
On the surface, we can uniquely (up to the rotation and the translation of the
space coordinates) choose a asymptotically Minkowski coordinate whose
time direction coincides with that of the asymptotically time-like Killing
vector.
By using this coordinate system, we can evaluate the mass uniquely from the
boundary values of the fields.

In the papers by Bilal, Kogan \cite{CII} and de Alwis \cite{CIII}, it was
considered the boundary value which corresponds to that on the surface of
the sphere in four-dimensional space time.
Since the two dimensional space is a line, its boundaries are two ends of
the line.
Bilal, Kogan and de Alwis have claimed that it is necessary to sum up these
two boundary values in order to get the mass.
We have questions about this point.
If we change the coordinate system, the time-slice \ie the line of
$t={\rm constant}$ and the positions of the end points of
the line are also changed.
Different coordinate choice gives different pair of two points which are
the boundaries of the space.
The direction and the scale of time at the two end points can be also changed
by the change of the coordinate.
This imply that the operation of the summing up two boundary values
does not have universal meaning.
Note that any observer lies on one of two end points.
If the mass which is measured by the observer depends on the boundary value
on another end point, which is space-likely disconnected with him, it seems
that the causality would be violated.
In this paper, we propose that the boundary value on the point where the
observer lives should be the mass of black hole which he observes.
Therefore the mass depends on which asymptotically flat region the observer
lives in.
More rigorously, we propose an observer-dependent mass as follows:
We choose a local Lorentz coordinate whose time-direction coincide with
the tangent of the observer's world line.
We choose a local Lorentz coordinate, where metric is given by
$g^{\mu\nu}={\rm diag}(-1,1)$, on the point where the observer lives.
We define the mass of the black hole by evaluating the boundary value in
this local Lorentz coordinate and we neglect the boundary value
on another point.\footnote{
If we define the mass by summing up two boundary value, the mass should be
positive semi-definite by using the argument of supersymmetry
\cite{BVII,BVIII}.
We could also show that the mass defined in this paper should be also bounded
from below if we could choose a coordinate system where an end of the space
is fixed to a point in the space-time.
If the mass measured by the fixed point is $m_0$, any mass should be greater
than $-m_0$ since the sum of the masses should be
positive due to the supersymmetry.
The result about the Bondi mass, which will be given later, might tell that
we cannot always choose the fixed point where the measured mass is finite.}

We now construct the Hamiltonian $H$ corresponding to the action
(\ref{exxxxivb}).
The conjugate momenta $\Pi_A\equiv 4\pi{\delta S \over \delta \dot A}$
($A=X,\bar Y,Z,\varphi,S,\bar T$, $x^\pm=t\pm r$ and
$\dot A\equiv \partial_t A$) are given by
\bea
\label{eeex}
\Pi_X&=&{k \over 2}\partial_t X \ , \nonumber \\
\Pi_{\bar Y}&=&{k \over 2}\Bigl\{-\sinh^2X\partial_t \bar Y
+\Bigl(\cosh^2 X-{1 \over 2}\Bigr)\partial_t Z\Bigr\} \ , \nonumber \\
\Pi_Z&=&{k \over 2}\Bigl\{\cosh^2 X\partial_t Z-\Bigl(\cosh^2X-{1 \over 2}
\Bigr)\partial_t\bar Y\Bigr\} \ , \nonumber \\
\Pi_\varphi&=&-{k \over 8}\partial_t\varphi \ , \nonumber \\
\Pi_S&=&-{k \over 8}\partial_t\bar T \ , \nonumber \\
\Pi_{\bar T}&=&-{k \over 8}\partial_t S \ .
\eea
In terms of $\Pi_A$ and $A'$ ($A'\equiv \partial_r A$), the Hamiltonian
density $T_{00}\equiv{1 \over 4}(T_{++}+T_{--}+2T_{+-})$ is given by
\bea
\label{eeexi}
T_{00}&=&k\Bigl\lbrack
{2 \over k^2}\Pi_X^2+{1 \over 2}X'^2
+{2 \over k^2}\cosh^{-2}X\Pi_Z^2
\ +{2 \over k}\Bigl(1-{1 \over 2}\cosh^{-2}X\Bigr)\Pi_Z\bar Y' \nonumber \\
&\ &+{1 \over 8}\cosh^{-2}X\,\bar Y'^2-{1 \over 8}\sinh^{-2}X\,Z'^2
\ -{2 \over k^2}\sinh^{-2}X\,\Pi_{\bar Y}^2 \nonumber \\
&\ & +{2 \over k}\sinh^{-2}X\Bigl(\cosh^2X-{1 \over 2}\Bigr)\Pi_{\bar Y}Z'
-{8 \over 8}\Pi_\varphi^2-{1 \over 8}\phi'^2 \nonumber \\
&\ &+{16 \over k^2}\Pi_S\Pi_{\bar T}+{1 \over 4}S'\bar T'+{1 \over 4}S''
\Bigr\lbrack
\nonumber \\
&\ &+{\alpha \over 2}\Bigl(\sinh X {\rm e}^{\bar Y+{\phi \over 2}}\Bigr)^{2l}
{\rm e}^{-2\bar T}+T_{00}^{{\rm matter}}+T_{00}^{{\rm ghost}}\ .
\eea
Then the infinitesimally small variation of the bulk Hamiltonian
$H_0=\int dr T_{00}$ with respect to $A$ and $\Pi_A$
($A=X,\bar Y,Z,\varphi,S,\bar T)$ is given by a bulk part $\delta H$,
which gives the equations of motions, and the surface term
$-\delta H_\partial$ (the minus sign comes from the later convienience.)
\be
\label{eeexii}
\delta H_0=\delta H -\delta H_\partial \ .
\ee
The explicit form of $-\delta H_\partial$ is given by,
\bea
\label{eeexiii}
-\delta H_\partial &=& k\Bigl\lbrack
X'\delta X+{2 \over k}\Bigl(1-{1 \over 2}\cosh^{-2}X\Bigr)\Pi_Z\delta \bar Y
+{1 \over 4}\cosh^{-2}X\,\bar Y'\delta\bar Y \nonumber \\
&\ & {1 \over 4}\sinh^{-2}X\,Z'\delta Z
+{2 \over k^2}\sinh^{-2}X\Bigl(\cosh^2 X-{1 \over 2}\Bigr)\Pi_{\bar Y}\delta Z
-{1 \over 4}\varphi'\delta\varphi \nonumber \\
&\ &+{1 \over 4}(\bar T'\delta S+S'\delta\bar T)+{1 \over 4}\delta S'
\Bigr\rbra
\Bigr|_{{\rm the\ end\ points\ of\ }t={\rm constant\ line}} \ .
\eea
By using the solution (\ref{eeeii}) of the equations of motion,
the equation (\ref{eeexiii}) can be rewritten by
\be
\label{eeexiv}
-\delta H_\partial = {k \over 4}(\bar T'\delta S+\delta S')
\Bigr|_{{\rm the\ end\ points\ of\ }t={\rm constant\ line}}\ ,
\ee
in the general coordinate system.
(Note that $\bar Y$ is a scalar but $\bar T$ is not scalar.)
Therefore the total Hamiltonian $H$ should be given by
\be
\label{eeexv}
H=H_0+H_\partial \ ,\ \ H_\partial= -{k \over 4}(\bar T'\Delta S+\Delta S')
\Bigr|_{{\rm the\ end\ points\ of\ }t={\rm constant\ line}}\ .
\ee
Here $\Delta S$ is a finite variation around some reference solution.
We need the above surface Hamiltonian $H_\partial$ in order to cancel
the surface term which comes from the variation of the bulk Hamiltonian $H_0$.
Since $T_{00}$ vanishes identically due to the constraints, $H_0$ also
vanishes and only $H_\partial$ can contribute to the value of the total
Hamiltonian $H$.
By assuming that an observer lives in one of two end points, we regard one
of the above two boundary values as a mass $M$ which the observer measures.
\be
\label{eeexvic}
M=-{k \over 4}(\bar T'\Delta S+\Delta S')
\Bigr|_{{\rm the\ end\ point\ where\ the\ observer\ lives}}\ .
\ee
In the neighborhood around the observer, we need to choose the local
Lorentz frame whose time direction is parallel to the tangent of the world
line of the observer.
Or equivalently, we can multiply $\sqrt{-g^{00}}$ to the expression in
(\ref{eeexvic}).
\be
\label{eeexvi}
M=-{k \over 4}\sqrt{-g^{00}}(\bar T'\Delta S+\Delta S')
\Bigr|_{{\rm the\ point\ where\ the\ observer\ lives}}\ .
\ee
The above expression (\ref{eeexvi}) can be used for any observer who does not
always live in an asymptotically flat region.
In this sense, we can define local mass by the equation (\ref{eeexvi}).

We now calculate the ADM mass of our model.
We choose the asymptotically Minkowski coordinates $\bar x^\pm$ for the
static solution (\ref{exxxiii}) ($b_+=b_-=0$) by
\begin{equation}
\label{exxxix}
\bar x^\pm=\pm{1 \over \lambda}\ln(\pm \lambda x^\pm) \ .
\end{equation}
and for the shock wave solution (\ref{exxxv}) ($a_\pm=b_\pm=0$) by
\begin{equation}
\label{exxxxi}
\bar x^+={1 \over \lambda}\ln(\lambda x^+) \ , \ \ \
\bar x^-=-{1 \over \lambda}\ln(-\lambda x^-
+{4m\lambda \over \Lambda k a}) \ .
\end{equation}
The ADM mass corresponds to the observer who lives in an asymptotically flat
region where $\bar x^++\bar x^-$ : fixed and
$\bar x^+-\bar x^-\rightarrow +\infty$ \ie $f\rightarrow +\infty$
(\ref{eeeiii}).
By using Eqs.(\ref{ei}) and (\ref{eeev}), we find the metric $g^{00}$ in
this region is given by
\bea
\label{eeexvii}
g^{00}&=&{1 \over 2}g^{+-}=-\e^{-2\rho} \nonumber \\
&\ &\rightarrow -{f\e^{-\lambda(\bar x^+-\bar x^-)} \over 2b
(2w^2)^{{1 \over a}}}
\nonumber \\
&\ &\rightarrow -{\Lambda \over 2b(2w^2)^{{1 \over a}}\lambda^2} \ .
\eea
Note that conformal mode $\rho$ transforms as
$\rho(x^\pm) \longrightarrow \rho(\bar x^\pm)
+{\lambda \over 2}(\bar x^+-\bar x^-)$ by the coordinate transformation
(\ref{exxxix}) or (\ref{exxxxi}).
In the following, we choose $\lambda$ so that $g^{00}\rightarrow -1$ \ie
\be
\label{eeexviii}
\lambda^2={\Lambda \over 2b(2w)^{{1 \over a}}}\ .
\ee
Then by using the mass formula (\ref{eeexvi}) and by choosing a reference
solution by
\begin{equation}
\label{eeexix}
-{ka \over 4}u_\pm^0(x^\pm)
=-{N \over 24}\{\lambda(\bar x^+-\bar x^-)-\ln\, \lambda^2\}  \ ,
\end{equation}
which gives a quantum analogue of the linear dilaton vacuum,
we find the ADM mass for the static solution
\be
\label{eeexx}
M={\lambda \over 2}(a_++a_-)\ ,
\ee
and for the shock wave solution,
\be
\label{eeexxi}
M=2m\e^{\lambda \bar x^+_0}\ ,
\ee
where $\bar x^+_0\equiv {1 \over \lambda}\ln(\lambda x^+_0)$.
Note that $\bar T$ in this coordinate system is given by
$\bar T=-{\lambda \over 2}(\bar x^+-\bar x^-)$.
The above values of masses (\ref{eeexx}) and (\ref{eeexxi}) do not vanish in
general and finite.

Besides $f\rightarrow +\infty$ region, there is another asymptotically flat
region where $f \rightarrow 0$ and we can also consider the mass measured by
an observer who lives in this region.
Due to Eq.(\ref{eeevi}), $g^{00}=-\e^{-2\rho}$ is given by,
\be
\label{eeexxii}
g^{00}\sim -f^{{1 \over a}+{1 \over 2}}\ .
\ee
Therefore we find the mass has the following form when $f\rightarrow 0$:
\be
\label{eeexxiii}
M\sim f^{{1 \over 2}\Bigl({1 \over a}+{1 \over 2}\Bigr)}\ln f \ .
\ee
Then we find the mass vanishes in the limit of $f\rightarrow 0$.

We now consider the Bondi mass. The Bondi mass is measured by the observer
who lives in the region where $\bar x^-$ fixed and
$\bar x^+ \rightarrow +\infty$.
Then the formula (\ref{eeexvi}) gives the following expression for the
shock wave solution:\footnote{
This expression is different from that in the previous paper \cite{CI},
where it was considered the first variations of the stress tensor with
respect to $X^\pm$. These fields, however, would be inadequate for the
calculation of the Bondi mass since they do not damp in the spacial
infinity.}
\be
\label{eeexxix}
M=2\lambda\Bigl\{{m \over \lambda}\e^{\lambda\bar x^+_0}
-{N \over 24}\ln\Bigl(1+{a \over \lambda}\e^{\lambda \bar x^-}\Bigr)
-{N \over 24}{\lambda \over 1+{a \over \lambda}\e^{\lambda \bar x^-}}
\Bigr\}\ .
\ee
The expression diverges when $\bar x^-\rightarrow +\infty$. The divergence,
however, would just tell that the Bondi mass is ill-defined and would
not give any other serious problem.
$\bar x^-\rightarrow +\infty$ means that the observer approaches to $f=0$
line.
In the following we show that the scalar curvature $R$ diverges
near $f=0$ line in the limit of $\bar x^+\rightarrow +\infty$ although
the curvature vanishes just on the $f=0$ line.
Since the space-time is strongly curved, the Bondi mass becomes ill-defined
in the limits $\bar x^+\rightarrow +\infty$ and $\bar x^-\rightarrow +\infty$.
The limit $\bar x^-\rightarrow +\infty$ does not commute with the limit
$\bar x^+\rightarrow +\infty$. The mass formula gives different answers
depending on different limiting procedures .

By defining $M$, $X$, $p$ and $q$ by
\bea
\label{eeeei}
&\ &M\equiv -{4mx^+_0 \over ka}\ , \ \ X\equiv -{4m \over \Lambda k a} \\
\label{eeeeii}
&\ &x^+=\e^{{1 \over 2}(p+q)}\ , \ \ x^-=-\e^{{1 \over 2}(p-q)}-X \ ,
\eea
$f$ in Eq.(\ref{eeeiii}) corresponding to the shock wave solution
(\ref{exxxv}) has the following form when $x^+>x^+_0$
\be
\label{eeeeiii}
f(p,q)=\Lambda\e^p+M-{N \over 24}\ln\Bigl\{\e^p\Bigl(
1+\e^{-{1 \over 2}(p-q)})\Bigr)\Bigr\} \ .
\ee
By putting $f=0$ and solving $q$ with respect to $p$, we obtain
\be
\label{eeeeiv}
q=q_0(p)\equiv -p+2\ln\Bigl\{{\e^{{24 \over N}(\Lambda\e^p+M)}-\e^p \over X}
\Bigr\}\ .
\ee
When $p$ is sufficiently large, $q$ becomes monotonously increasing
function with respect to $p$ and behaves $q\sim {48 \over N}(\Lambda\e^p+M)$.
Therefore in the limit $p\rightarrow +\infty$ on the $f=0$ line, $x^\pm$ go
to $x^+\rightarrow +\infty$ and $x^-\rightarrow -X$ \ie
$\bar x^\pm\rightarrow +\infty$.

We now calculate the scalar curvature
$R=8\e^{-2\rho}\partial_+\partial_-\rho=-8\e^{-2\rho-p}
(\partial_p^2-\partial_q^2)\rho$. Here $\rho$ is given in Eq.(\ref{eeev}).
By defining
\be
\label{eeeev}
P\equiv p-p_0\ , \ \ Q\equiv -q+q_0(p_0) \ ,
\ee
we consider the limit $p_0\gg |P|, |Q|$ and calculate the scalar curvature
$R$ on the line where $Q=0$. On the $Q=0$ line, $P\geq 0$.
When $p_0\gg |P|, |Q|$, $f$ has the following form
\be
\label{eeeevi}
f=\Lambda\e^{p_0}(\e^p-1)-{N \over 48}(P+Q)+{\it o}(1)\ ,
\ee
which tells that the derivative with respect $Q$ can be neglected compared
with the derivative with respect to $P$. Then we obtain
\be
\label{eeeevii}
R\sim -8\e^{-2\rho -(P+p_0)}(\partial_p^2f\rho'+(\partial_pf)^2\rho'')\ .
\ee
Here $\rho'={d\rho \over df}$. (See Eq.(\ref{eeev}).)

When $0\leq P\ll \e^{-p_0}$, $f$ is small : $f\ll 1$.
By using Eq.(\ref{eeevi}), we find
\be
\label{eeeeviia}
R\sim -4\Lambda^{{1 \over a}+{1 \over 2}}({1 \over a}+{1 \over 2})
\e^{\Bigl({1 \over a}-{1 \over 2}\Bigr)p_0}P^{{1 \over a}-{3 \over 2}}\ .
\ee
When $22<c_{{\rm matter}}<24$, $R$ vanishes when $P=0$ but the derivative
of $R$ with respect to $P$ exponentially increases, due to the factor
$\e^{\Bigl({1 \over a}-{1 \over 2}\Bigr)p_0}$, as $p_0$ increases.

On the other hand, when $p$ is finite and $p_0\gg 1$, $f$ is large $f\gg 1$.
Then, by using Eq.(\ref{eeev}), we find that $\rho$ behaves as
\be
\label{eeeeviii}
\rho\sim-{1 \over 2}\ln f+{\cal O}(1)\ ,
\ee
and the scalar curvature $R$ is given by
\be
\label{eeeeix}
R\sim -{4\Lambda \over \e^P-1 }\ .
\ee
%The scalar curvature is nothing but that of the classical black hole if we
%identify $\e^P=x^+x^-$.
The expression (\ref{eeeeix}) tells that $R$ diverges in the limit
$P\rightarrow 0$, where the expression (\ref{eeeeix}) breaks down and the
expression (\ref{eeeeviii}) becomes valid. By combining these results
Eq.(\ref{eeeeix}) and Eq.(\ref{eeeeviii}), we find that $R$ has a peak
$|R|\sim \e^{p_0}$
when $P\sim \e^{-p_0}$ and the scalar curvature in the neighborhood of
the $f=0$ line diverges in the limit of $p_0\rightarrow +\infty$
($x^+\rightarrow +\infty$, $x^-\rightarrow -X$).
At present we do not fully understand the meaning of this curvature
singularity.

Instead of defining Bondi mass, we consider the mass measured by an observer
who is depicted in Fig.4. He lives in an asymptotically flat
$f\rightarrow +\infty$ region at first and he approaches to $f=0$ line.
For example, we assume the world line of the observer is given by
$r\equiv {1 \over 2}(\bar x^+-\bar x^-)=r_0$ (constant) and the constant
$r_0$ is sufficiently large.
When he lives in the asymptotically flat region, the mass measured by him is
given by Eq.(\ref{eeexxi}).
As he approaches to $f=0$ line, the space-time becomes to have a curvature
and the mass measured by him becomes to vary and he would consider he is
observing the Hawking radiation.
After infinite proper time, he reaches to the $f=0$ line, where the measured
mass vanishes and he would consider the black hole has evaporated completely.
Of course, since the space-time is curved in the neighborhood of the $f=0$
line, the mass measured there has not universal meaning. The masses,
however, measured by at first (in the asymptotically flat region) and at last
(on the $f=0$ line) are universal. Therefore a class of observers who live in
the asymptotically flat region at first and whose world lines cross the $f=0$
line in finite $x^+$ (but after infinite proper time) observes the black hole
evaporation.

\section{Summary and Discussion}

In this paper, we have analyzed a quantum two dimensional dilaton gravity
model, which is described by $SL(2,R)/U(1)$ gauged Wess-Zumino-Witten model
deformed by $(1,1)$ operator.  It has been shown that the curvature
singularity does not appear when the central charge $c_{\rm matter}$ of the
matter fields is given by $22<c_{\rm matter}<24$. When $22<c_{\rm matter}<24$,
the matter  shock  waves,  whose  energy  momentum  tensors  are  given  by
$T_{\rm matter} \propto \delta(x^+ - x^+_0)$, create a kind of wormholes,
\ie causally disconnected regions. Most of the quantum informations in
past  null  infinity  are  lost  in  future  null infinity but the lost
informations would be carried by the wormholes.

Recently Hawking and Hayward \cite{OO}
have pointed that there might be a close
connection between wormholes and the formation and evaporation of black
holes.
In their scenario, informations falling into black holes pass into another
universe through the wormholes.
It is interesting to consider the relation between their works and the present
one.

We have also discussed  about the problem of defining the mass of quantum
black holes.  On the basis of the argument by Regge and Teitelboim, we find
the ADM mass  measured  by  the  observer  who  lives in one of asymptotically
flat regions is finite and does not vanish in general.
Instead of the Bondi mass, we have considered a mass measured by a class of
observers who live in an asymptotically flat region at first and approaches to
$f=0$ line.
They observe the change of mass, \ie the Hawking radiation.
The measured mass vanishes finally and the black hole evaporates completely.
Therefore the total Hawking radiation is positive even when $N<24$.

\vskip 1cm

\noindent
{\bf Acknowledgement}

We would like to acknowledge A. Sugamoto and K. Odaka for for discussions.
We are also grateful to other members of particle physics group in
Ochanomizu University for their hospitality.

%%%%%%

\pagebreak

\newpage

\centerline{\bf Figure Captions}

\vskip 1cm

\noindent
{\bf Fig.1}
The structure of the space-time corresponding to a static solution
is depicted. $p$ and $q$ are defined by
$x^\pm\equiv\pm \e^{{1 \over 2}(p\pm q)}$. The asymptotically flat region
corresponding to $f\rightarrow +\infty$ is depicted like a cylinder.
And the asymptotically flat region corresponding to
$f\rightarrow +0$ is depicted like a plane.

\vskip 0.5cm

\noindent
{\bf Fig.2}
Wormhole creation by a matter shock wave.
The region I is causally disconnected with asymtotically Minkowski
region II.

\vskip 0.5cm

\noindent
{\bf Fig.3}
Worm holes created by two matter shock waves.
The two shock wave solution creates two (Fig.3a) or two (Fig.3b) wormhole.

\vskip 0.5cm

\noindent
{\bf Fig.4}
The world line of the observer is depicted. His world line is given by
$r\equiv {1 \over 2}(\bar x^+-\bar x^-)=r_0$ (constant).
He lives in an asymptotically flat $f\rightarrow +\infty$ region at first and
he approaches to $f=0$ line.

\end{document}